\documentclass{article}
\title{Quantum Cryptography with Squeezed States}
\author{Mark Hillery \\ Department of Physics and 
Astronomy \\ Hunter College of the City University of
New York \\ 695 Park Avenue \\ New York, NY 10021 }
\begin{document}
\maketitle
\begin{abstract}
A quantum key distribution scheme based on the use of squeezed
states is presented.  The states are squeezed in one of two
field quadrature components, and the value of the squeezed
component is used to encode a character from an alphabet.
The uncertainty relation between quadrature components 
prevents an eavesdropper from determining both with enough
precision to determine the character being sent.  Losses
degrade the performance of this scheme, but it is possible
to use phase sensitive amplifiers to boost the signal
and partially compensate for their effect.
\end{abstract}
\section{Introduction}
Quantum cryptography provides a means of sending a secure message,
and it does this by allowing one to establish a secure key.  In all 
but the simplest codes, what is sent is not only the coded message, 
but also a key which tells the receiver how to decode the message.
The coded message can be sent through a public channel, but the
key must be sent through a secure one.  Quantum mechanics allows
one to construct a channel in which the presence of an eavesdropper
can be detected \cite{qcrypt1}-\cite{qcrypt3}.  
The key can be sent through this
channel, and if no eavesdropping is found, the key will be secure.
Working quantum cryptographic systems have been constructed in
several laboratories \cite{benn3}-\cite{hughes}.

The quantum cryptographic schemes proposed so far have all
involved the transmission of single particles.  For example,
in one scheme, single photons are sent down an optical fiber, 
and information is carried by the polarization
of the photons.  Experimental implementations of this method
use weak coherent pulses rather than single photons.
Losses limit the distances over which this
method can be used; if the fiber is too long, the probability
of the photon emerging from the fiber without being absorbed 
is small.  Amplifiers cannot be used to boost the signal, 
because they destroy the quantum coherence which is essential
for the method to work.  A second approach, which also suffers
from this limitation, uses weak, overlapping coherent states
\cite{benn1}-\cite{hutt}.  In this case, the information is
encoded in the phase of the coherent state.

One possible way around the limitation imposed by losses 
is to use pulses 
consisting of more than one photon.  Care, however, is 
required, because the eavesdropper may siphon off enough 
of the pulse to learn what information it is carrying, 
but send the rest of the pulse on its way.  It has been 
shown by Ralph that multiphoton pulses
in coherent states are vulnerable to this kind of attack
\cite{ralph}.  It is necessary to use pulses for which this 
kind of eavesdropping will not work.

Recently Ralph has presented a method of using squeezed
states for quantum cryptography \cite{ralph}.  In this 
scheme sequences of symbols are impressed on two squeezed
beams by Alice.  The beams are then mixed at a beam 
splitter, and the mixed beams, along with a local oscillator
for each, is sent to Bob.  A random phase delay is introduced
into one of the beams and its local oscillator in order to
destroy the phase coherence between the two beams.  Bob,
by using both beams and their local oscillator signals,
can, by using homodyne detections, recover one of the two 
sequences but not both.  An
eavesdropper is in the same position as Bob, but she
does not know which sequence Bob will read, and if she
uses a capture-resend strategy, every time she guesses 
incorrectly, she will introduce detectable
errors.  The squeezing prevents her from gaining useful
information by splitting off parts of the beams.

Here we shall investigate a different scheme based on 
squeezed light.  Alice sends displaced squeezed vacuum
states to Bob which are squeezed in one of two orthogonal
field quadrature components.  Bob chooses at random which
of the components to measure.  The security of this method 
of transmission is a result of
the uncertainty relation for field quadrature components.
The effect of loss is examined, and it is found that
its effect can be partially compensated by using 
degenerate parametric amplifiers to boost the signal.
This method should be secure against the capture-resend
strategy and a strategy which employs a beam splitter to 
sample part of the signal.  In the latter case, it is the
vacuum noise which presents a major problem for the 
eavesdropper.

\section{Procedure}
A single-mode classical field is characterized by a complex
amplitude, or equivalently, by its real and imaginary
parts, which we shall designate by $x_{1}$ and $x_{2}$,
respectively.  Quantum mechanically, the complex amplitude
corresponds to the mode annihilation operator, $a$, and the
real and imaginary parts to the operators $X_{1}$ and $X_{2}$,
respectively, where
\begin{equation}
X_{1}=\frac{1}{2}(a^{\dagger}+a) \hspace{1cm}
X_{2}=\frac{i}{2}(a^{\dagger}-a) .
\end{equation}
These operators do not commute and obey the uncertainty
relation ($\hbar =1$)
\begin{equation}
\Delta X_{1} \Delta X_{2}\geq \frac{1}{4} .
\end{equation}
This uncertainty relation implies that $X_{1}$ and $X_{2}$
cannot both be defined to arbitrarily high accuracy for a given
quantum state.  Is is this fact which will form the basis of
our quantum cryptography system.

It is often useful to represent quantum states in a phase
space whose axes are $x_{1}$ and $x_{2}$.  The state is 
pictured as a point surrounded by an error box.  The point
is located at $x_{1}=\langle X_{1}\rangle$ and $x_{2}=
\langle X_{2}\rangle$, and the error box represents the
fluctuations of the amplitude about its mean value.  For
a coherent state, the error box is a circle of radius $1/2$,
while for a minimum uncertainty squeezed state, it is 
an ellipse whose minor axis is parallel to
the direction of the
squeezing.  The area of the ellipse is the same as that of 
the circle.  Its shape, however, allows us to have one of 
the variables, say $X_{1}$, very precisely defined, while
the other, $X_{2}$, is very poorly defined.

In explaining how this can be used to send a message, we
invoke the usual cast of characters which appears in
discussions of quantum cryptography, Alice, Bob, and Eve.
Alice wants to establish a key with Bob, and Eve wants to
intercept it without being detected.  Alice and Bob use
the following method to set up a shared key.  The $x_{1}$
and $x_{2}$ axes are divided up into bins of size $\delta$,
where $\delta < 1/2$.  Each bin corresponds, by previous 
agreement, to a symbol in an alphabet.  The key will consist 
of a sequence of symbols from this alphabet.  For example,
a not very efficient choice of alphabet would be would 
be the symbols $0$ and $1$, and every other bin could 
represent $0$ while the intervening ones represent $1$.
A more efficient choice would be to use a larger
alphabet.  

Alice now sends to Bob one of two kinds of squeezed
states.  The first kind can be represented by an ellipse
which is centered on the $x_{1}$ axis and is squeezed 
in the $x_{1}$ direction to a width considerably less
than $\delta$.  This type of state has very well-defined
$x_{1}$ value but a poorly defined $x_{2}$ value.  The
second kind is represented by an ellipse which is 
centered on the $x_{2}$ axis and is squeezed in the 
$x_{2}$ direction, also to a width considerably less than
$\delta$.  This state has a well-defined $x_{2}$ value
but a poorly defined $x_{1}$ value.  We shall call the
first kind of state an $x_{1}$ state and the second kind
an $x_{2}$ state.

The number of bins on each axis depends on the length
of the ellipses.  Let $\delta x_{maj}$ be the length of
the major axis of the ellipses representing the
$x_{1}$ and $x_{2}$ states (it is assumed to be the 
same for both).  On the $x_{1}$ axis, the bins run from 
$-\delta x_{maj}/2$ to $\delta x_{maj}/2$, and have 
the same range on the $x_{2}$ axis. This means that the
collection of all $x_{1}$ states, each centered in a
particular bin on the $x_{1}$ axis, covers the same
region of phase space as does the collection of
$x_{2}$ states.  Therefore, an eavesdropper cannot
determine whether a state is an $x_{1}$ or an $x_{2}$
state from the result of a single measurement of
$X_{1}$ or $X_{2}$.

Alice now decides at random whether to send an $x_{1}$
or an $x_{2}$ state, and Bob decides, also at random,
whether to measure $X_{1}$ or $X_{2}$.  This measurement
can be performed by using homodyne detection.  Alice
and Bob then communicate with each other via a public
channel.  For each state which Alice sent, she tells 
Bob what kind of state, $x_{1}$ or $x_{2}$, it was, and
Bob tells Alice what kind of measurement he made.
Alice does not tell Bob the $X_{1}$ or $X_{2}$ values of
the states she sent, and Bob does not tell Alice the 
results of his measurements.  After this public
communication, Alice and Bob keep the results
for which Bob made a measurement corresponding to the
state which Alice sent, e.\ g.\ when Alice sent an $x_{1}$
state and Bob measured $X_{1}$, and discard the others.
For each of these transmissions, Bob knows the value of
the variable, $x_{1}$ or $x_{2}$, which Alice sent so
they both know which bin the state falls in.  They then 
assign to this transmission the alphabet symbol 
corresponding to this bin.  The result is a sequence of 
symbols which can be used as a key.

Why is this key secure?  In order to know which bin a 
given state falls into, Eve must be able to determine either
$x_{1}$ or $x_{2}$ to an accuracy of at least $\delta$.
The problem is, she does not know which measurement to make, 
and she is forbidden by the uncertainty principle from
measuring both to the necessary accuracy.  She must
choose to measure $X_{1}$ or $X_{2}$ if she wishes to
determine the key symbol, and if she chooses the
wrong one, she gains no information and disturbs the
message.  This disturbance can be detected by Alice and
Bob.  They can compare a subset of the transmissions
for which they should agree.  If they find errors, 
i.\ e.\ if they find they do not agree on some of these
symbols, they can conclude there was an eavesdropper
present.

Let us now see how small $\delta$ needs to be, and
how much squeezing we need.  For a
squeezed vacuum state, squeezed in the $x_{1}$ direction,
the probability distribution for the observable $X_{1}$
is given by (see Appendix A)
\begin{equation}
\label{probx1}
p(x_{1})=\langle x_{1}|\rho_{sqvac}|x_{1}\rangle =
\frac{1}{\sqrt{\pi v}}e^{-x_{1}^{2}/v} ,
\end{equation}
where $\rho_{sqvac}$ is the density matrix corresponding
to the squeezed vacuum state, $v=(1/2)e^{-2r}$, and 
$r\geq 0$ is the squeezing parameter ($r=0$ corresponds
to the vacuum state which is not squeezed).  The
probability, $p_{\delta}$, that $x_{1}$ lies in the 
interval $[-\delta /2,\delta /2]$ is
\begin{equation}
\label{pdelta}
p_{\delta}=\frac{2}{\sqrt{\pi v}}\int_{0}^{\delta /2}
dx_{1}e^{-x_{1}^{2}/v} = {\rm erf}\left( \frac{\delta}
{2\sqrt{v}}\right) ,
\end{equation}
where erf$(x)$ is the error function and is given by
\begin{equation}
{\rm erf}(x)=\frac{2}{\sqrt{\pi}}\int_{0}^{x}dt 
e^{-t^{2}} .
\end{equation}
Suppose we want the probability of making an error,
i.\ e.\ the probability of finding $x_{1}$ outside
the inverval $[-\delta /2,\delta /2]$, to be less
than $10^{-3}$.  We find that $1-{\rm erf}(2.51)
=3.9\times 10^{-4}$ \cite{abram}, so that if
\begin{equation}
\delta \geq 5.02\sqrt{v} ,
\end{equation}
then the error probability will be less than $10^{-3}$.
If we choose $\delta$ to be $1/8$, then we can choose
$v=6.2\times 10^{-4}$ which implies that the squeezing 
parameter is 3.3.  Because the number of photons in a
squeezed vacuum is $\langle a^{\dagger}a\rangle =
\sinh^{2}r$, the number of photons in this state is
$200$. The width of the state in the
$x_{2}$ direction is just $e^{r}/2$ which is, with
our choices, $14$, and this implies that in that 
direction the state has substantial overlap with
approximately $110$ bins.  This means that if we
measure the state in the $x_{1}$ direction we can
determine with very good probability which bin it 
lies in, but if it is measured in the $x_{2}$
direction, the result is essentially random.

\section{Effect of Losses}
As the light travels down a fiber it will experience 
losses and this will degrade the squeezing.  These 
losses can be described by the master equation
\begin{equation}
\label{mast}
\frac{d\rho}{dt}=\frac{\gamma}{2}(2a\rho a^{\dagger}
-a^{\dagger}a\rho-\rho a^{\dagger}a) .
\end{equation}
Here $\rho$ is the density matrix of the field and
$\gamma$ is the loss rate.  In order to find the 
density matrix at the end of the fiber, one solves
this equation for $\rho (t)$ and sets $t=T=L/c$, 
where $L$ is the length of the fiber.  There are 
a number of ways to solve this equation one of which is 
discussed Appendix B.  The result for $p(x_{1})$
at the output of the fiber is again given by
Eq. (\ref{probx1}), but now
\begin{equation}
v=\frac{1}{2}[(1-e^{-\gamma T})+e^{-\gamma T}e^{-2r}] .
\end{equation}
The probability of finding $x_{1}$ within a bin of
size $\delta$ is still given by Eq. (\ref{pdelta}), 
but with the new value of $v$.  For a given value
of $p_{\delta}$, or, equivalently, a given error
probability, this relation gives a bound on the size
of the loss which can be tolerated.  Note that 
even if the initial squeezing were infinite (a
physical impossibility, because this would require
infinite energy), the size of the acceptable loss
is finite, and, in fact, quite small.  For example,
for an error probability of less than $10^{-3}$, 
with the same value of $\delta$ as above, we
can again choose $v=6.2\times 10^{-4}$ which gives
us that that $\gamma T <1.2\times 10^{-3}$ or
for a fiber of length one kilometer, a maximum
loss of $1.2\times 10^{-6}$ per meter.

It is possible to use a degenerate parametric
amplifier to partially compensate for the effect
of losses.  In order to see how this works, we shall
compare the action of a fiber of length $L=Tc$ to
that of two fibers of length $L/2$ with a degenerate
parametric amplifier, with gain $G>1$, between
them to boost the signal.  We shall consider what
happens to an $x_{1}$ state, the effect on an
$x_{2}$ state is similar.

In the case of the fiber of length $L$ we have
that
\begin{eqnarray}
\langle X_{1}(T)\rangle & = & e^{-\gamma T/2}
\langle X_{1}(0)\rangle , \nonumber \\
\Delta X_{1}(T)^{2} & = & e^{-\gamma T}\Delta X_{1}
(0)^{2}+\frac{1}{4}(1-e^{-\gamma T}) .
\end{eqnarray}
Because we wish to find which bin the initial state
is in, we define a variable $\xi = e^{\gamma T/2}
X_{1}(T)$ which has the property that 
\begin{eqnarray}
\langle \xi\rangle & = & \langle X_{1}(0)\rangle , 
\nonumber \\
\Delta \xi & = & [\Delta X_{1}(0)^{2}+\frac{1}{4}
(e^{\gamma T}-1)]^{1/2} .
\end{eqnarray}
A measurement of $\xi$ will tell us with high
probability which bin the original state was in if
$\Delta \xi$ is considerably smaller than $\delta$,
or
\begin{equation}
\Delta\xi <s\delta ,
\end{equation}
where $s<1$, and its actual size is determined by the 
probability of error which we can tolerate.  In
order to satisfy this condition the loss must be such
that
\begin{equation}
e^{\gamma T}-1 \cong \gamma T <4[(s\delta )^{2}-
\Delta X_{1}(0)^{2}] .
\end{equation}

Now let us look at the case with the amplifier.
After the first fiber we have
\begin{eqnarray}
\label{L/2}
\langle X_{1}(T/2)\rangle & = & e^{-\gamma T/4}
\langle X_{1}(0)\rangle , \nonumber \\
\Delta X_{1}(T/2)^{2} & = & e^{-\gamma T/2}
\Delta X_{1}(0)^{2}+\frac{1}{4}(1-e^{-\gamma T/2}) .
\end{eqnarray}
The amplifier can be set to amplify either $X_{1}$
or $X_{2}$; if it amplifies $X_{1}$, then $X_{1}
\rightarrow GX_{1}$ and $X_{2}\rightarrow (1/G)
X_{2}$, while if it is set to amplify $X_{2}$, then
$X_{1}\rightarrow (1/G)X_{1}$ and $X_{2}\rightarrow
GX_{2}$.  Let us suppose that it is set to amplify
$X_{1}$ which has the effect of multiplying the 
right-hand side of the first of Eqs. (\ref{L/2}) by
$G$ and the second by $G^{2}$.  Finally, after the
second fiber we have
\begin{eqnarray}
\langle X_{1}(T)\rangle & = & Ge^{-\gamma T/2}
\langle X_{1}(0)\rangle , \nonumber \\
\Delta X_{1}(T)^{2} & = & G^{2}e^{-\gamma T}
\Delta X_{1}(0)^{2}+\frac{1}{4}G^{2}e^{-\gamma T/2}
(1-e^{-\gamma T/2}) \nonumber \\
 & & +\frac{1}{4}(1-e^{-\gamma T/2}) .
\end{eqnarray}
Again we are interested in the value of $X_{1}$ at the
beginning of the first fiber, so we define 
\begin{equation}
\xi_{1}=\frac{1}{G} e^{\gamma T/2}X_{1}(T) ,
\end{equation}
which implies that
\begin{eqnarray}
\label{right}
\langle \xi_{1}\rangle & = & \langle X_{1}(0)\rangle , 
\nonumber \\
\Delta \xi_{1} & = & [\Delta X_{1}(0)^{2}+\frac{1}{4}
(e^{\gamma T/2}-1)+\frac{1}{4G^{2}}e^{\gamma T}(1-
e^{-\gamma T/2})]^{1/2} .
\end{eqnarray}
In the limit of large gain, the requirement on the loss 
so that $\Delta \xi_{1}<s\delta$ is that (for $\gamma
T \ll 1)$
\begin{equation}
\label{lossreq}
\gamma T<8[(s\delta )^{2}-\Delta X_{1}(0)^{2}] ,
\end{equation} 
a less stringent requirement by a factor of two over 
the single long fiber.  Therefore, the amplifier, by
boosting the signal, has reduced the effect of the 
losses.

It remains to be seen what happens if the amplifier
is set to amplify $X_{2}$, and an $x_{1}$ state is
sent.  In that case we define the variable
\begin{equation}
\xi_{2} = Ge^{\gamma T/2}X_{1}(T) ,
\end{equation}
which has the following properties:
\begin{eqnarray}
\label{wrong}
\langle \xi_{2}\rangle & = & \langle X_{1}(0)\rangle , 
\nonumber \\
\Delta \xi_{2} & = & [\Delta X_{1}(0)^{2}+\frac{1}{4}
(e^{\gamma T/2}-1)+\frac{1}{4}G^{2}e^{\gamma T}(1-
e^{-\gamma T/2})]^{1/2} .
\end{eqnarray}
In the high gain limit the requirement on the losses
is
\begin{equation}
\gamma T<8\frac{(s\delta)^{2}}{G^{2}} .
\end{equation}
This is a much more stringent requirement on the 
losses that that imposed by the single long fiber.

This analysis suggests that Alice and Bob should use 
the following protocol if a fiber with amplifiers is
to be used.  Alice decides at random whether to send
an $x_{1}$ or an $x_{2}$ state and independently
decides, also at random, whether the amplifiers
should amplify $X_{1}$ or $X_{2}$.  She then
sends the state, and Bob measures either $X_{1}$
or $X_{2}$, with this choice being again random.
If Alice and Bob make the same choice, and the 
amplifiers are set the same way, e.\ g.\ both
decide to measure $X_{1}$ and the amplifiers
also amplify $X_{1}$, then they can use that
transmission, otherwise they discard it.  

If the amplifier settings are secure, then this
procedure can be simplified and Alice can set the
amplifiers in accord with the state she sends.  On
the other hand, if they are not, and Eve knows which
quadrature component is being amplified, then the
random setting is necessary.  There is, in addition,
a limit on the gain of the amplifiers which follows from
the fact that Alice and Bob must be able to use the 
cases in which Alice's preparation and Bob's measurement
agree, but the amplifier is set incorrectly, to detect
the presence of the eavesdropper.

If Alice and Bob can only use the cases in which
everything agrees, preparation, amplifier setting, and
measurement, to detect Eve, then she has a successful
eavesdropping strategy which cannot be detected.  Eve
simply determines which quadrature is being amplified,
measures that quadrature, and and sends a state which
agrees with the result of her measurement on to Bob.
Using this strategy Eve will only make incorrect
measurements when the amplifier is set to amplify
the wrong quadrature component, in which case that
transmission will be discarded anyway.  Thus, it is
essential that Alice and Bob be able to examine the
transmissions for which their preparation and
measurement agree, but the amplifier is set 
incorrectly, and see the effect of Eve's measurement.

Let us now see what kinds of restrictions this
requirement imposes.  In the case in which the
amplifier is set properly, we want the third
term in the brackets in the second of Eqs.
(\ref{right}) to be much smaller than the other
two.  This implies that we need
\begin{eqnarray}
\Delta X_{1}(0)^{2} & \gg & \frac{\gamma T}{G^{2}} 
\nonumber \\
\gamma T & \gg & \frac{\gamma T}{G^{2}} .
\end{eqnarray}
The second of these inequalities implies that we
simply need $G^{2} \gg 1$, and if we assume that
$\Delta X_{1}(0) \sim s\delta$, and in addition
assume that the requirement in Eq. (\ref{lossreq})
is obeyed, then the first of the above inequalities
also reduces to $G^{2} \gg 1$.  Now let us see
what happens when the amplifier is set incorrectly.
With no intervention by Eve, we have that $\Delta
\xi_{2}\sim G$.  If, however, Eve makes an incorrect
measurement, then the uncertainty in the result of
her measurement will be of order $1/(s\delta )$.
This uncertainty will be reflected in the state 
she sends to Bob, and consequently he will find
$\Delta \xi_{2}\sim 1/(s\delta )$.  If $G$ is
chosen so that
\begin{equation}
G^{2}\gg 1 \hspace{1cm} G\ll \frac{1}{s\delta} ,
\end{equation}
then even by examining the cases in which the 
amplifier is set incorrectly, Alice and Bob can,
by comparing their results, tell whether an
eavesdropper was present.

\section{Eavesdropping}
Let us now consider two possible methods of 
eavesdropping on this system.  The first is just
the capture-resend strategy, and the second involves
using a beam splitter to split off part of the 
signal, and performing measurements on that part
to gain information about the signal.  In both
cases, we shall find that the intervention of the
eavesdropper is detectable.

In the capture-resend strategy, Eve measures the entire
signal, and then, on the basis of her measurement 
result, prepares a second state which she sends on to
Bob.  Her problem is that she does not know whether
she should measure $X_{1}$ or $X_{2}$, and, if
$\delta$ is chosen small enough, she will introduce
errors if she chooses incorrectly.
 
If $\delta$ is chosen too large, in particular, larger 
than $1/2$, Eve has a straightforward eavesdropping 
strategy.  She can measure both $X_{1}$ and $X_{2}$ to
an accuracy of approximately $1/2$.  This can be
accomplished either by splitting the signal into two
parts using a 50-50 beam splitter and measuring $X_{1}$
at one output and $X_{2}$ at the other \cite{leon}, or
by amplifying the signal, so that it becomes essentially 
classical, and performing measurements on it 
\cite{schleich}.  After performing these measurements,
she sends a coherent state to Bob which is centered
on the results of her measurement.  That is, if she
obtained results $x_{1}^{(m)}$ and $x_{2}^{(m)}$, the 
coherent state she sends can be visualized as a circle 
of radius $1/2$ in the $x_{1}$-$x_{2}$ plane with its 
center at the point $(x_{1}^{(m)}, x_{2}^{(m)})$.
When Alice announces which kind of state she sent, Eve
knows which of her results, $ x_{1}^{(m)}$ or 
$x_{2}^{(m)}$, to use, and Alice and Bob will not be
able to detect the eavesdropping.  This is because the 
coherent state which Eve sent to Bob will have the
correct value of $x_{1}$ for an $x_{1}$ state and the
correct value of $x_{2}$ for an $x_{2}$ state, and
both with sufficiently high accuracy that Bob will
assign the result of his measurement to the correct
bin.  Therefore, it is necessary to choose $\delta$
smaller than $1/2$ in order to foil this strategy.

If $\delta$ has been chosen considerably smaller
than $1/2$ the above strategy no longer works,
because Eve can no longer determine both $X_{1}$
and $X_{2}$ to the desired accuracy, i.\ e.\ 
$\delta$.  She will have to choose which to 
measure, and after making the measurement, will
send to Bob a state squeezed in the direction
which she chose centered on the result of her
measurement.  If, however, she made the wrong
choice, the state she sends to Bob will be
squeezed in the wrong direction and centered
on the wrong point.  This will introduce errors,
which, by comparing a subset of the results on
which they agree, i.\ e.\ Bob's measurement
corresponded to the kind of state which Alice 
sent, Alice and Bob can detect the eavesdropping.

Let us now suppose that $\delta$ has been chosen
sufficently small, and that instead of measuring
the entire signal, Eve uses a beam splitter to
sample a part of it.  She sends on to Bob the
part of the signal which is transmitted through
the beam splitter and performs measurements on the
part which is reflected.  We would like to see how
much she can learn, and how much she disturbs
the signal state.  We shall call the two modes
which the beam splitter couples modes $1$ and $2$,
and the signal will go into the input port for
mode $1$ and the vacuum into the input for mode
$2$.  The relation between the input and output
operators is \cite{yurke}
\begin{equation}
\label{beam}
\left( \begin{array}{c} a^{(out)}_{1} \\
a^{(out)}_{2} \end{array}\right) = U^{-1}
\left( \begin{array}{c} a^{(in)}_{1} \\
a^{(in)}_{2} \end{array}\right) U
= \left(
\begin{array}{cc}\sqrt{T} & \sqrt{R} \\
-\sqrt{R} & \sqrt{T} \end{array} \right)
\left( \begin{array}{c} a^{(in)}_{1} \\
a^{(in)}_{2} \end{array}\right) ,
\end{equation}
where $U$ is the unitary operator which implements
the beam-splitter transformation, and
$R$ and $T$ are the reflection and transmission
coefficients, respectively.  These have to be chosen
in such a way that we minimize the disturbance to
the signal state, but, nevertheless, gain some
information about it.  We shall suppose that
Alice has sent an $x_{1}$ state, and see
what happens both in the case Eve makes an $X_{1}$ 
measurement and in the case she makes an $X_{2}$
measurement.

Now suppose that Alice has sent an $x_{1}$ state 
centered on $x_{1}=s$ with
squeezing parameter $r$, and let $\sigma =e^{-r}$.  
Eve inserts the beam splitter and measures
$X_{1}$ at the mode $2$ output port.  We shall denote 
by $X_{jk}$ the operator
$X_{j}$ for the $k$th mode, where both $j$ and $k$ 
can be either $1$ or $2$.  We shall also drop the
superscript $(in)$ on all $in$ operators with the 
understanding that operators without a superscript
are $in$ operators.  From Eq. (\ref{beam}) we have
that for initially uncorrelated modes
\begin{eqnarray}
\langle X_{12}^{(out)}\rangle & = & \sqrt{T}\langle 
X_{12}\rangle -\sqrt{R}\langle X_{11}\rangle 
\nonumber \\
(\Delta X_{12}^{(out)})^{2} & =& T(\Delta X_{12})^{2}
+R(\Delta X_{11})^{2} ,
\end{eqnarray}
or for our input state,
\begin{eqnarray}
\label{x12out}
\langle X_{12}^{(out)}\rangle & = & -\sqrt{R}s
\nonumber \\
(\Delta X_{12}^{(out)})^{2} & =& \frac{1}{4}(T
+R \sigma^{2}) .
\end{eqnarray}
This last equation tells us about the information
gain from Eve's measurement.  In order to learn
about Alice's state with some degreee of accuracy,
$\Delta X_{12}^{(out)}$ cannot be too large which,
in turn, implies that the reflection coefficient 
cannot be too small.  If it is, the noise from the 
vacuum state obscures the information carried by
the signal state.  In particular, if we define
$\xi_{12}=-X_{12}^{(out)}/\sqrt{R}$, then we have
\begin{eqnarray}
\langle \xi_{12}\rangle & = & s \nonumber \\
(\Delta\xi_{12})^{2} & = & \frac{1}{4}\left( \frac
{T}{R}+\sigma^{2}\right) .
\end{eqnarray}
From this equation it is clear that if we want
to determine $x_{11}$ with an accuracy of order
$\delta$, then we must have $\sqrt{T/R}$
of order $\delta$.

Let us now see what is the effect of Eve's
measurement on the transmitted signal state.  In
the $x_{1}$ representation the initial wave function
of the system is (see Appendix A)
\begin{equation}
|\Psi\rangle = \psi_{x_{1}}(x_{11})\phi_{vac}(x_{12}) ,
\end{equation}
where
\begin{equation}
\label{befx1}
\psi_{x_{1}}(x_{11})=
\left( \frac{2}{\pi \sigma^{2}}\right)^{1/4}
e^{-[(x_{11}-s)/\sigma]^{2}} ,
\end{equation}
is the wave function of the $x_{1}$ state,
\begin{equation}
\phi_{vac}(x_{12}) = \left( \frac{2}{\pi}\right)^{1/4}
e^{-x_{12}^{2}} ,
\end{equation}
is the wave function of the vacuum state, and $x_{11}$
is the $x_{1}$ coordinate for mode $1$ and $x_{12}$ is
the $x_{1}$ coordinate for mode $2$.  After the
beam splitter the wave function is (see Appendix A)
\begin{equation}
U|\Psi\rangle = \left( \frac{2}{\pi \sigma}\right)^{1/2}
e^{-[(\sqrt{T}x_{11}-\sqrt{R}x_{12}-s)/\sigma]^{2}}
e^{-(\sqrt{R}x_{11}+\sqrt{T}x_{12})^{2}} .
\end{equation}
If Eve now measures $X_{12}$ and obtains the result
$y$, the wave function becomes a product of a
wave function in mode $1$ and an ``eigenstate'' of
$X_{12}$ with eigenvalue $y$ in mode $2$.  The mode
$1$ wave function, $\psi_{y}(x_{11})$ is then
\begin{equation}
\label{aftx1}
\psi_{y}(x_{11})= N\exp \left[ -\left(\frac{T}{\sigma^{2}}
+R\right)\left( x_{11}-\frac{\sqrt{T}(s+\sqrt{R}
(1-\sigma^{2})y)}{T+R\sigma^{2}}\right) \right] ,
\end{equation}
where $N$ is a normalization constant.  This is the
wave function which will be sent on to Bob.
Comparing Eqs.(\ref{befx1}) and (\ref{aftx1}) we
see that there are two effects of the measurement on
the wave function.  First, the center of the 
Gaussian has shifted, and, second, its width has 
changed.  We want both of these changes to be
small.

As a result of the beam splitter and measurement,
the width has changed as follows:
\begin{equation}
\sigma\rightarrow \frac{\sigma}{(T+\sigma^{2}R)^{1/2}} .
\end{equation}
In order for this change to be small, it is necessary
that $T$ be of order one.  The shift in the center of 
the Gaussian is given by
\begin{equation}
s\rightarrow \frac{\sqrt{T}(s+\sqrt{R}(1-\sigma^{2})y)}
{T+\sigma^{2}R}
\end{equation}
Note that if $T=1$ the center suffers no shift while
if $T=0$ it is shifted all the way to zero.  This
clearly implies that in order to produce a small
change we want $T$ to be close to one.  Assuming
this to be the case, which means that $R\ll 1$, 
and also that $\sigma$ is small, we find that the 
shift, $\Delta s$, is given by
\begin{equation}
\Delta s=-\frac{1}{2} sR+y\sqrt{R} .
\end{equation}
Now $y$ is random, but will typically be of order
$-\sqrt{R}s$ (see Eq. (\ref{x12out})) so that if
the shift is to be less than $\delta$ in magnitude,
we must have
\begin{equation} 
sR<\delta .
\end{equation}
However, $s$ can be anywhere between $-1/\sigma$ and 
$1/\sigma$, so that we need to require that
\begin{equation}
R< \sigma \delta ,
\end{equation} 
which is a very stringent requirement.

Comparing the requirements for information gain
and small disturbance, we see that they are 
incompatible.  Information gain requires a small
transmission coefficient, while a small disturbance
requires a transmission coefficient close to one,
and there is no overlap in the permitted ranges.
Therefore, using this method, if Eve diverts 
enough light to gain useful information, she will
also produce a detectable disturbance.  The real
problem for Eve is the vacuum noise.  If she
samples only a small part of the signal, in order
to minimize the disturbance, what little signal
she sees is swamped by vacuum noise.

\section{Conclusion}
A method for using squeezed states to perform
quantum key distribution has been presented.  It
relies on the uncertainty relation for field
quadrature components for its security.  As with
other methods, it is adversely affected by losses,
but it is possible in this case to use amplifiers
to reduce their effect.  We have shown that this
method is secure against several eavesdropping
strategies, but have not presented a general proof
of its security.

Squeezed states are an example of a nonclassical 
field state, that is a state whose photodetection
properties cannot be simulated by a classical
stochastic field.  They have proven to be useful
for teleportation \cite{braun1}-\cite{braun3} and 
now for quantum cryptography as well.  This leads
one to ask whether other kinds of nonclasical states
could prove useful in quantum information, and 
suggests that this should be a fruitful line of 
research.

\section*{Acknowledgments}
This work was supported by the National Science
Foundation under grant PHY-9970507 and by a grant 
from PSC-CUNY.

\section*{Appendix A}
Here we present several facts about squeezed states
and beam splitters which are needed in the rest of
the paper.  Squeezed states are discussed in several
recent textbooks on quantum optics, and these provide
a good background for the subject \cite{scully,mandel}.

A squeezed vacuum state is obtained from the vacuum
by applying the squeeze operator
\begin{equation}
|\Phi\rangle = S(z)|0\rangle ,
\end{equation}
where
\begin{equation}
S(z)=\exp [(z(a^{\dagger})^{2}-z^{\ast}a^{2})/2] .
\end{equation}
Setting $z=re^{i\phi}$ we have that
\begin{eqnarray}
\label{sqeez}
S(z)^{-1}aS(z) & = & a\cosh r + a^{\dagger}e^{i\phi}
\sinh r \nonumber  \\
S(z)aS(z)^{-1}& = & a\cosh r - a^{\dagger}e^{i\phi}
\sinh r .
\end{eqnarray}

We can use these relations to find an explicit
expression for $|\Phi\rangle$ in the $x_{1}$
representation, i.\ e.\ the representation of
$|\Phi\rangle$ given by $\langle x_{1}|\Phi\rangle$,
where $|x_{1}\rangle$ is an eigenstate of $X_{1}$.
In this representation
\begin{equation}
X_{1}\rightarrow x_{1} \hspace{1cm} X_{2}\rightarrow
-\frac{i}{2}\frac{d}{dx_{1}} ,
\end{equation}
which implies that 
\begin{equation}
\label{annx1}
a\rightarrow x_{1}+\frac{1}{2}\frac{d}{dx_{1}}
\hspace{1cm} a^{\dagger}\rightarrow x_{1}
-\frac{1}{2}\frac{d}{dx_{1}} .
\end{equation}
The state $|\Phi\rangle$ satisfies the equation
\begin{equation}
S(z)aS(z)^{-1}|\Phi\rangle = 0,
\end{equation}
and using Eqs. (\ref{sqeez}) and (\ref{annx1}) to
put this in the $x_{1}$ representation we find
\begin{eqnarray}
0 & = & \left[ (\cosh r-e^{i\phi}\sinh r)x_{1} 
\right. \nonumber \\
 & & \left. +\frac{1}{2}(\cosh r+e^{i\phi}\sinh r)
\frac{d}{dx_{1}}\right] \langle x_{1}|\Phi\rangle .
\end{eqnarray}
This equation is easily solved, and in the case
$\phi = \pi$, which corresponds to squeezing in the
$x_{1}$ direction, we find, after normalization, that
\begin{equation}
\langle x_{1}|\Phi\rangle = \left(\frac{2}{\pi e^{-2r}}
\right)^{1/4}\exp [-(x_{1}/e^{-r})^{2}] .
\end{equation}
In order to obtain the wave function of a squeezed
vacuum state which has been shifted by $s$ in the 
$x_{1}$ direction, we simply replace $x_{1}$ by
$x_{1}-s$ in the above equation.

Next we would like to see how wave functions in
the $x_{1}$ representation are transformed
under the action of a
beam splitter.  Let our initial state be
$\langle x_{11},x_{12}|\Psi\rangle = \Psi_{in}
(x_{11},x_{12})$, then our task is to find
\begin{equation}
\Psi_{out}(x_{11},x_{12}) 
= \langle x_{11},x_{12}|U\Psi\rangle , 
\end{equation}
where $U$ is the beam-splitter
transformation given in Eq. (\ref{beam}).  We first
note that $U^{-1}|x_{11},x_{12}\rangle$ is an eigenstate 
of $U^{-1}X_{11}U$ with eigenvalue $x_{11}$, and 
of $U^{-1}X_{12}U$ with eigenvalue $x_{12}$.  From
Eq. (\ref{beam}) we have
\begin{eqnarray}
U^{-1}X_{11}U & = & \sqrt{T}X_{11}+\sqrt{R}X_{12}
\nonumber \\
U^{-1}X_{12}U & = & -\sqrt{R}X_{11}+\sqrt{T}X_{12} ,
\end{eqnarray}
so that 
\begin{equation}
U^{-1}|x_{11},x_{12}\rangle = |\sqrt{T}x_{11}-\sqrt{R}
x_{12},\sqrt{R}x_{11}+\sqrt{T}x_{12}\rangle .
\end{equation}
Therefore, we have that
\begin{equation}
\Psi_{out}(x_{11},x_{12})=\Psi_{in}(\sqrt{T}x_{11}
-\sqrt{R}x_{12},\sqrt{R}x_{11}+\sqrt{T}x_{12}) .
\end{equation}

\section*{Appendix B}
We want to find the solution to the master equation, 
Eq. (\ref{mast}).  There are a number of ways of doing
this, most of which involve turning the operator 
equation into a c-number equation.  We shall use 
the master equation to derive an equation for the 
symmetrically-ordered field characteristic function,
which for a single-mode field is given by
\begin{equation}
\chi (\xi )={\rm Tr}(D(\xi )\rho ) ,
\end{equation}
where $D(\xi )=\exp (\xi a^{\dagger}-\xi^{\ast}a)$.
using the relations 
\begin{eqnarray}
\frac{\partial\chi}{\partial\xi}-\frac{1}{2}\xi^{\ast}
\chi & = & {\rm Tr}(D(\xi )a^{\dagger}\rho )
\nonumber \\
\frac{\partial\chi}{\partial\xi}+\frac{1}{2}\xi^{\ast}
\chi & = & {\rm Tr}(D(\xi )\rho a^{\dagger})
\nonumber \\
\frac{\partial\chi}{\partial\xi^{\ast}}+\frac{1}{2}\xi
\chi & = & -{\rm Tr}(D(\xi )a\rho )
\nonumber \\
\frac{\partial\chi}{\partial\xi^{\ast}}-\frac{1}{2}\xi
\chi & = & -{\rm Tr}(D(\xi )\rho a) ,
\end{eqnarray}
we can transform the master equation into a partial
differential equation for $\chi (\xi)$
\begin{equation}
\frac{\partial\chi}{\partial t}=-\frac{\gamma}{2}
\left( \xi\frac{\partial\chi}{\partial\xi}+\xi^{\ast}
\frac{\partial\chi}{\partial\xi^{\ast}}+|\xi |^{2}
\chi \right) .
\end{equation}
Defining $\chi^{\prime}(\xi )=\exp (|\xi|^{2}/2)
\chi (\xi )$, we find that $\chi^{\prime}(\xi )$
satisfies
\begin{equation}
\frac{\partial\chi^{\prime}}{\partial t}=
-\frac{\gamma}{2} \left( \xi\frac{\partial\chi^{\prime}}{\partial\xi}+\xi^{\ast}
\frac{\partial\chi^{\prime}}{\partial\xi^{\ast}}
\right) ,
\end{equation}
whose solution is given by
\begin{equation}
\chi^{\prime}(\xi, t)=\chi^{\prime}(e^{-\gamma t/2}
\xi ,0) .
\end{equation}
This implies that
\begin{equation}
\label{chi}
\chi (\xi, t)=\exp [-(1-e^{-\gamma t})|\xi |^{2}/2]
\chi (e^{-\gamma t/2}\xi ,0) . 
\end{equation}

Our next task is to relate the characteristic
function to the $x_{1}$ distribution of the
density matrix.  If we let $q$ and $p$ be the real
and imaginary parts of $\xi$, $\xi =q+ip$, then
\begin{eqnarray}
\label{chix1}
\chi(\xi ) & = & e^{-iqp}{\rm Tr}(e^{2ipX_{1}}
e^{-2iqX_{2}}\rho ) ,\nonumber \\
 & = & e^{-iqp}\int dx_{1}e^{2ipx_{1}}\langle x_{1}|
e^{-2iqX_{2}}\rho |x_{1}\rangle .
\end{eqnarray}
Setting $q=0$ we see that $\langle x_{1}|\rho |x_{1}
\rangle$ is just the Fourier transform of $\chi (p)$,
so
\begin{equation}
\label{x1chi}
\langle x_{1} |\rho |x_{1}\rangle = \frac{1}{\pi}
\int dp e^{-2ix_{1}p}\chi (p) .
\end{equation}

In order to find $\langle x_{1}|\rho (t)|x_{1}
\rangle$ for an initial squeezed vacuum state, we
first use Eq. (\ref{chix1}) and the results of
Appendix A to find $\chi (p)$ at $t=0$.  We then
use Eq. (\ref{chi}) to find $\chi (p,t)$, and 
finally, Eq. (\ref{x1chi}) to find $\langle x_{1}
|\rho (t)|x_{1}\rangle $.

\bibliographystyle{unsrt}

\end{document}